\documentclass[twocolumn,showpacs,10pt]{revtex4-1}%
\usepackage{amsmath,amssymb,graphicx}
\usepackage{amsmath}
\usepackage{amsfonts}
\usepackage{amssymb}
\usepackage{graphicx}%
\providecommand{\U}[1]{\protect\rule{.1in}{.1in}}
\begin{document}
\title{Quantum defogging: temporal photon number fluctuation correlation in time-variant fog scattering medium}
\author{Deyang Duan, Yuge Li, Yunjie Xia}
\affiliation{School of Physics and Physical Engineering, Qufu Normal University, Qufu 273165, China}
\begin{abstract}
The conventional McCartney model simplifies fog as a scattering medium with
space-time invariance, as the time-variant nature of fog is a pure noise for
classical optical imaging. In this letter, an opposite finding to traditional
idea is reported. The time parameter is incorporated into the McCartney model
to account for photon number fluctuation introduced by time-variant fog. We
demonstrated that the randomness of ambient photons in the time domain results
in the absence of a stable correlation, while the scattering photons are the
opposite. This difference can be measured by photon number fluctuation correlation
when two conditions are met. A defogging image is reconstructed from the target’s
information carried by scattering light. Thus, the noise introduced by time-variant
fog is eliminated by itself. Distinguishable images can be obtained even when the
target is indistinguishable by conventional cameras, providing a prerequisite
for subsequent high-level computer vision tasks.

\end{abstract}
\maketitle

Images captured outdoors usually suffer from noticeable degradation due to
light scattering and absorption, especially in extreme weather, such as fog
and haze. The restoration of such hazy images has become an important problem
in many computer vision applications such as visual surveillance, remote
sensing, and autonomous transportation. Early methods tried to estimate a
transmission map with physical priors and then restore the corresponding image
via a scattering model, e.g., the dark channel prior [1]. However, these
physical priors are not always reliable, leading to inaccurate transmission
estimates and unsatisfactory dehazing results. With the advances achieved in
deep learning, many methods based on convolutional neural networks and
generative adversarial networks have been proposed to overcome the drawbacks
of using physical priors [2-9]. These techniques are more efficient and
outperform traditional prior-based algorithms. In common cases, large
quantities of paired hazy$/$clean images are necessary for model training.
However, it is almost impossible to obtain these image pairs from the real
world, and most learning-based methods resort to training on synthetic data.
When a hazy image is completely indistinguishable, this pairing
relationship cannot be established at all.

Mie established a scattering model and revealed the reason for image
degradation in scattering media. Then, McCartney proposed a simplified
scattering model [10]. Two assumptions are contained in this model: (i) the
distance between the object plane and image plane is limited; and (ii) the
scattering medium is uniformly distributed on the light path and has
space-time invariance. The second item is reasonable in conventional optical
cameras since the exposure time is sufficiently short. Thus, it can be
approximately considered that the physical properties of scattering media are
space-time invariant in a single measurement event. However, this assumption
is no longer valid when continuously shooting or when a single measurement
time is long because the photon number fluctuation introduced by the
time-variant fog is significant in the time domain. Whether classical optical
imaging or ghost imaging, photon number
fluctuation is always considered a purely negative factor
because it leads to obvious contrast and visibility degradation [1-9,11-19].
Consequently, increasing attention has been given to recovering clean images
from a single hazy image [2-9].
\begin{figure}[ptbh]
\centering
\fbox{\includegraphics[width=1\linewidth]{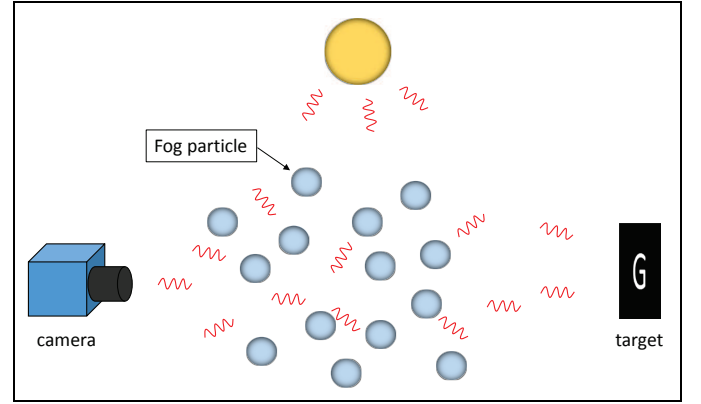}}\caption{Setup of the
McCartney model.}%
\label{fig:false-color}%
\end{figure}

Recall that the scattering model proposed McCartney, including a scattering
light model and an ambient light model, has the following expression [10]:%
\begin{equation}
I\left(  d,\lambda\right)  =I_{0}\left(  \lambda\right)  e^{-\beta\left(
\lambda\right)  d}+I\left(  \infty,\lambda\right)  \left(  1-e^{-\beta\left(
\lambda\right)  d}\right)  .
\end{equation}
Here, $I\left(  d,\lambda\right)  $ represents the light measured by the
camera, which corresponds to the image output by the camera. $d$ and $\lambda$
represent the distance from the object plane to the image plane and the
wavelength, respectively. $I_{0}\left(  \lambda\right)  $ represents the light
emitted or reflected by the object, which corresponds to the image without
fog. $\beta\left(  \lambda\right)  $ represents the scattering coefficient.
$I\left(  \infty,\lambda\right)  $ represents the ambient light, which
corresponds to the noise. Notably, the fog image is a linear superposition of
the scattering light (signal) and ambient light (noise).

The conceptual arrangement of our imaging technique based on a McCartney model
is illustrated in Fig. 1. In the scattering light model, the irradiance of a
single measurement event passed through a volume element with a unit
cross-sectional area, a thickness of $dx$ and a period of time $dt$ can be
expressed as%
\begin{equation}
\frac{dI\left(  x,\lambda,t\right)  }{I\left(  x,\lambda,t\right)  }%
=-\beta\left(  \lambda,t\right)  dxdt,
\end{equation}
where $\beta\left(  \lambda,t\right)  $ represents the time-dependent
scattering coefficient. $x$ presents the vertical coordinate. By integrating
Eq. 2, we obtain%
\begin{equation}
I\left(  d,\lambda,t\right)  =\int_{t_{1}}^{t_{2}}I_{0}\left(  \lambda
,t\right)  e^{-\beta\left(  \lambda,t\right)  d}dt,
\end{equation}
where $\beta\left(  \lambda,t\right)  d$ denotes the optical path, and $\Delta
t=t_{2}-t_{1}$ represents the integration time. We adopt a statistical view of
light: a point source of radiation randomly emits photons in all possible
directions [20]. Thus, the radiation measured during $\Delta t$ is the result
of the superposition of a large number of photons or subfields. We have
$\int_{\Delta t}I_{0}\left(  \lambda,t\right)  dt=\sum_{m=1}^{n}E_{m}\left(
\lambda,t\right)  $ with $E_{m}\left(  \lambda,t\right)  $ representing the
$m$th photon or subfield, $n\propto\Delta t$. The measured light intensity,
which is proportional to the number of photons, corresponds to the theoretical
expectation [20], i.e.,%
\begin{equation}
I\left(  d,\lambda,t\right)  =\sum_{m=1}^{n}E_{m}\left(  \lambda,t\right)
\int_{t_{1}}^{t_{2}}e^{-\beta\left(  \lambda,t\right)  d}dt,
\end{equation}

In the ambient light model, the luminous intensity in volume $dv$ in the
measurement direction is%
\begin{equation}
dE\left(  x,\lambda,t\right)  =kx^{2}\beta\left(  \lambda,t\right)  d\omega
dxdt,
\end{equation}
where $dv=$ $d\omega x^{2}dx$, in which $d\omega$ is the solid angle from the
lens to the target, and $k$ is a function related to the scattering type.
Thus, the residual radiance obtained after scattering by the medium can be
expressed as%
\begin{equation}
dL\left(  x,\lambda,t\right)  =\frac{dE\left(  x,\lambda,t\right)
e^{-\beta\left(  \lambda,t\right)  x}}{x^{2}}.
\end{equation}
We obtain the light intensity, i.e.,%
\begin{equation}
dI\left(  x,\lambda,t\right)  =\frac{dL\left(  x,\lambda,t\right)  }{d\omega}.
\end{equation}
Substituting Eq. 5 and Eq. 6 into Eq. 7, we obtain%
\begin{equation}
dI\left(  x,\lambda,t\right)  =k\beta\left(  \lambda,t\right)  e^{-\beta
\left(  \lambda,t\right)  x}dxdt.
\end{equation}
The light intensity detected by the camera is obtained by integrating the
above equation, i.e.,%
\begin{equation}
I\left(  d,\lambda,t\right)  =k\sum_{m=1}^{n}E_{m}^{^{\prime}}\left(
\lambda,t\right)  \int_{t_{1}}^{t_{2}}\left(  1-e^{-\beta\left(
\lambda,t\right)  d}\right)  dt.
\end{equation}
Practically, the scattering light and ambient light are simultaneously
measured by the camera. Consequently, the fog image can be described as a
linear superposition of the scattering light and ambient light, i.e.,%
\begin{align}
I\left(  d,\lambda,t\right)   &  =\sum_{m=1}^{n}E_{m}\left(  \lambda,t\right)
\int_{t_{1}}^{t_{2}}e^{-\beta\left(  \lambda,t\right)  d}dt\nonumber\\
&  +\sum_{m=1}^{n}E_{m}^{^{\prime}}\left(  \lambda,t\right)  \int_{t_{1}%
}^{t_{2}}\left(  1-e^{-\beta\left(  \lambda,t\right)  d}\right)  dt\\
&  =S\left(  d,\lambda,t\right)  +A\left(  d,\lambda,t\right)  .\nonumber
\end{align}
Notably, both the scattering light and ambient light have photon number
fluctuations, whether in one measurement event or in different measurement events.

A set of experimental results are used to demonstrate the above theory. The
experimental setup is shown in Fig. 1. A quasi-monochromatic cosine LED with
$\lambda=$528$\pm$10 nm illuminates an object (letter G), and then the
reflected light is detected by a conventional camera (the imaging source is
DFK23U618) through a time-variant fog medium with a length of 0.6 m. Fog is
produced by 100\% water. Fig. 2 shows that the photon number fluctuation is
proportional to the integration time required for one measurement event.
Moreover, the photon number fluctuation phenomenon is also significant for
different measurement events. The quantitative results are shown in Fig. 3.
The structural similarity (SSIM) and peak signal-to-noise
ratio (PSNR) are used to present the photon number fluctuation results.
\begin{figure}[ptbh]
\centering
\fbox{\includegraphics[width=1\linewidth]{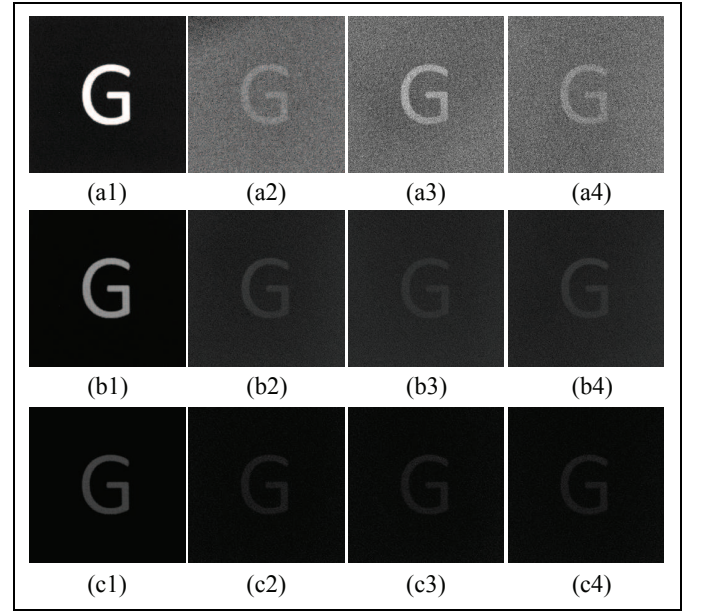}}\caption{The images output
by a conventional optical camera without fog (first column) and with fog
(columns 2 to 4). The integration times are 1/30 second (top row), 1/50 second
(middle row) and 1/150 second (bottom row). }%
\label{fig:false-color}%
\end{figure}\begin{figure}[ptbh]
\centering
\fbox{\includegraphics[width=1\linewidth]{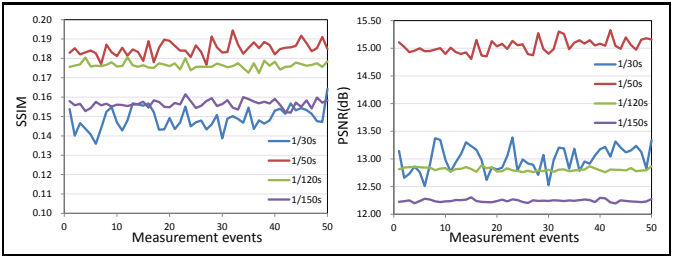}}\caption{SSIM and PSNR
curves produced for different integration times and different measurement
events.}%
\label{fig:false-color}%
\end{figure}

The photon number correlations between any two adjacent measurement events
along the time axis are expressed as shown in [21]:%
\begin{equation}
C\left(  d,\lambda,t\right)  =\left\langle I_{1}\left(  d,\lambda
,t_{1}\right)  I_{2}\left(  d,\lambda,t_{2}\right)  \right\rangle .
\end{equation}
To determine the result of Eq. 11, we consider the coherence time $\tau_{c}$
of the light field. Previous works showed that when $\tau_{c}>\Delta\tau$,
where $\Delta\tau$ represents the time interval between two measurement
events, we have that $\left\langle I_{1}I_{2}\right\rangle \neq0$ [23-25].
Thus, Eq. 11 can be expressed as
\begin{align}
C\left(  d,\lambda,t\right)   &  =\left\langle S_{1}\left(  d,\lambda
,t_{1}\right)  S_{2}\left(  d,\lambda,t_{2}\right)  \right\rangle \nonumber\\
&  +\left\langle A_{1}\left(  d,\lambda,t_{1}\right)  A_{2}\left(
d,\lambda,t_{2}\right)  \right\rangle .
\end{align}
The above result is still a linear superposition of signal and noise.
If $\tau_{c}<\Delta\tau$, we
have $\left\langle I_{1}I_{2}\right\rangle =0$ [22-24]. Thus, no image is
obtained. Zerom \emph{et al}. showed that $\left\langle I_{1}I_{2}%
\right\rangle \neq0$ can be guaranteed by accumulating the intensity
fluctuations of the light field at a certain time [24]. Nevertheless, the
result is the same as Eq. 12.

Considering the photon number correlation between two adjacent measurement
events of ambient light along the time axis, i.e.,%
\begin{equation}
\left\langle A_{1}\left(  d,\lambda,t_{1}\right)  A_{2}\left(  d,\lambda
,t_{2}\right)  \right\rangle =\left\vert G^{(1)}\left(  t_{1},t_{2}\right)
\right\vert ^{2},%
\end{equation}
where $G^{(1)}$ is usually defined as the first-order coherence function [20].
If $\tau_{c}<\Delta\tau$, $G^{(1)}\left(  t_{1},t_{2}\right)
=0$. Consequently, this type of light does not have stable photon number
correlation. Actually, the ambient photons
randomly come from all possible directions and point sources. Moreover, due to the photon number accumulation, we have
$\left\langle A_{1}A_{2}\right\rangle \rightarrow0$ but $\left\langle
A_{1}A_{2}\right\rangle \neq0$.

The photon number correlation of scattering
light can be expressed as%
\begin{equation}
\left\langle S_{1}\left(  d,\lambda,t_{1}\right)  S_{2}\left(  d,\lambda
,t_{2}\right)  \right\rangle =\left\vert G^{^{\prime}(1)}\left(  t_{1}%
,t_{2}\right)  \right\vert ^{2}.%
\end{equation}
Here, every point on the object can
be seen as a point source that emits photon. These photons pass through the
scattering medium and are focused onto an image plane by an imaging lens
defined by the Gaussian thin lens equation $1/s_{o}+1/s_{i}=1/f,$ where
$s_{o}$ is the distance between the object and the imaging lens, $s_{i}$ the
distance between the imaging lens and the image plane, and $f$ the focal
length of the imaging lens. Thus, scattering photons follow the point-point
mapping relationship between the object plane and image plane. Due to
the scattering of the medium, we have%
\begin{equation}
0<\frac{\left\langle S_{1}\left(  d,\lambda,t_{1}\right)  S_{2}\left(
d,\lambda,t_{2}\right)  \right\rangle }{\left\langle S_{1}\left(
d,\lambda,t_{1}\right)  \right\rangle \left\langle S_{2}\left(  d,\lambda
,t_{2}\right)  \right\rangle }<1.
\end{equation}
Consequently, scattering light exhibits stable photon number correlation,
even if photon number fluctuations still exist.

According to above analysis, if the measurement events meet the following
conditions, i.e., (i) the measurement events can present spatial photon number
fluctuations in time domain (i.e., $\left\langle I_{1}\right\rangle
\neq\left\langle I_{2}\right\rangle $), and (ii) the time interval between two
adjacent measurement events is larger than the coherence time of light field
(i.e., $\left\langle S_{1}S_{2}\right\rangle \neq0$, $\left\langle A_{1}%
A_{2}\right\rangle =0$). Eq. 12 can be expressed as
\begin{equation}
C\left(  d,\lambda,t\right)  =\left\langle S_{1}\left(  d,\lambda
,t_{1}\right)  S_{2}\left(  d,\lambda,t_{2}\right)  \right\rangle .
\end{equation}
Eq. 16 presents a fog-free image with low brightness. In practice, Eq. 12 is
rewritten as%
\begin{align}
C\left(  d,\lambda,t\right)   &  =\left\langle S_{1}\left(  d,\lambda
,t_{1}\right)  S_{2}\left(  d,\lambda,t_{2}\right)  \right\rangle \nonumber\\
&  +\min\left\langle A_{1}\left(  d,\lambda,t_{1}\right)  A_{2}\left(
d,\lambda,t_{2}\right)  \right\rangle ,
\end{align}
where $\min\left\langle A_{1}A_{2}\right\rangle $ represents $\left\vert
\min\left\langle A_{1}A_{2}\right\rangle \right\vert \ll\left\vert
\left\langle S_{1}S_{2}\right\rangle \right\vert $. Consequently, a
high-quality defogging image can be reconstructed.
\begin{figure}[ptbh]
\centering
\fbox{\includegraphics[width=1\linewidth]{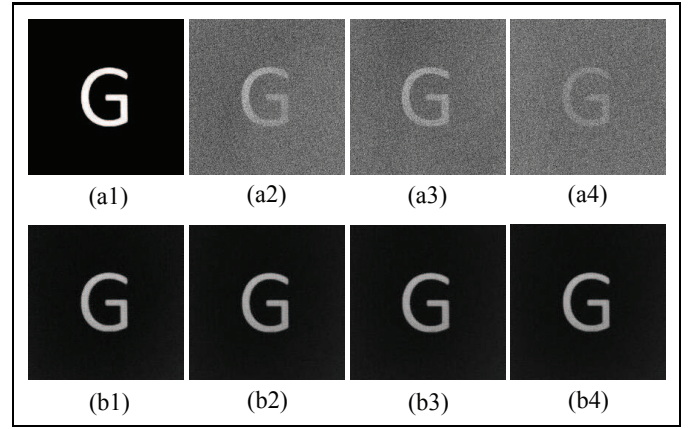}}\caption{A set of fog
images and reconstructed images. The integration time $\Delta t= 1/30$ second.
The time interval between two adjacent measurement events is $\Delta\tau=1$
second. The reconstructed image is obtained by measuring 20 fog images. (a1)
The target image, (a2)-(a4) three fog images, and (b1)-(b4) a set of
reconstructed images.}%
\label{fig:false-color}%
\end{figure}

Figs. 4 and 5 are used to demonstrate Eq. 17. The experimental data are
processed by the photon number correlation algorithm [25-28].
The results (Fig. 4) show that the influence of fog is almost completely
eliminated. Additionally, the image brightness is slightly lower than that of
the image without fog because the fluctuating photons are eliminated. The
quantitative results (Fig. 5) show that the SSIM and PSNR values of the
reconstructed images are much higher than those those of fog images, but
SSIM$\neq1$. The compared video is shown in supplemental material. Moreover, the
quantitative comparison between our method and the defogging generative
adversarial network is shown in Fig. 5 (for details, see supplemental material).
\begin{figure}[ptbh]
\centering
\fbox{\includegraphics[width=1\linewidth]{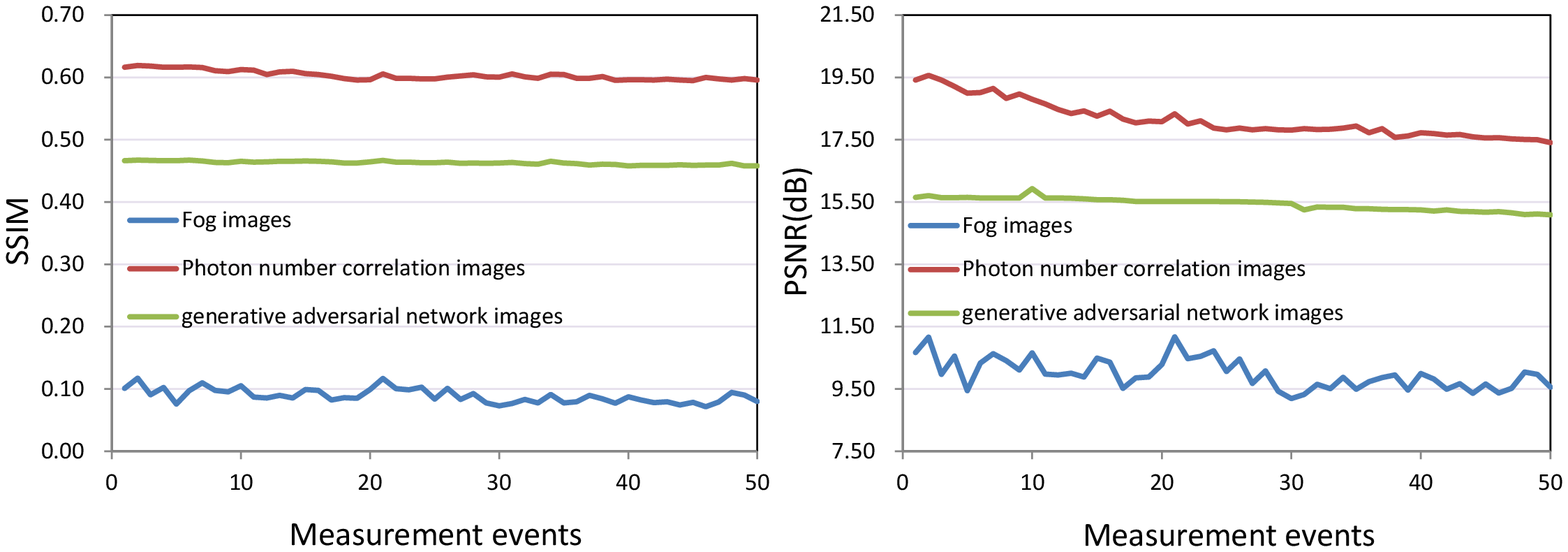}}\caption{SSIM and PSNR
curves of fog images and reconstructed images via photon number correlation
and images via generative adversarial network.}%
\label{fig:false-color}%
\end{figure}

Actually, the physical essence of measuring the photon number
correlation in the time domain is to eliminate
fluctuating photons through the fluctuating photons themselves. Consequently, a photon
number fluctuation correlation algorithm is introduced to eliminate the
fluctuating photon based on photon number fluctuation-fluctuation correlations
[20,29,30]. First, the software records the registration time of each measurement
event of a CCD camera element along the time axis. Then, the software calculates the average number of items per set:
$\overline{p_{1}}$ and $\overline{p_{2}}$. Subsequently, the software
classifies the values in each window as positive and negative fluctuations
based on $\overline{p_{1}}$ and $\overline{p_{2}}$.%
\begin{align}
\Delta p_{j\alpha}^{\left(  +\right)  }  &  =\left\{
\begin{array}
[c]{c}%
p_{j\alpha}-\overline{p_{j}},if,p_{j\alpha}>\overline{p_{j}}\\
0,otherwise
\end{array}
\right.  ,\nonumber\\
\Delta p_{j\alpha}^{\left(  -\right)  }  &  =\left\{
\begin{array}
[c]{c}%
p_{j\alpha}-\overline{p_{j}},if,p_{j\alpha}<\overline{p_{j}}\\
0,otherwise
\end{array}
\right.  ,
\end{align}
where $j=1,2$ labels two sets and $\alpha=1,2,...N$ labels the ath
measurement. event. $N$ is the total size of each set. We define the following
quantities:%
\begin{align}
\left(  \Delta p_{1}\Delta p_{2}\right)  _{\alpha}^{\left(  ++\right)  }  &
=\left\vert \left(  \overline{p_{1}}-\Delta p_{1\alpha}^{(+)}\right)  \left(
\overline{p_{2}}-\Delta p_{2\alpha}^{(+)}\right)  \right\vert ,\nonumber\\
\left(  \Delta p_{1}\Delta p_{2}\right)  _{\alpha}^{\left(  --\right)  }  &
=\left\vert \left(  \overline{p_{1}}-\Delta p_{1\alpha}^{(-)}\right)  \left(
\overline{p_{2}}-\Delta p_{2\alpha}^{(-)}\right)  \right\vert ,\nonumber\\
\left(  \Delta p_{1}\Delta p_{2}\right)  _{\alpha}^{\left(  +-\right)  }  &
=\left\vert \left(  \overline{p_{1}}-\Delta p_{1\alpha}^{(+)}\right)  \left(
\overline{p_{2}}-\Delta p_{2\alpha}^{(-)}\right)  \right\vert ,\nonumber\\
\left(  \Delta p_{1}\Delta p_{2}\right)  _{\alpha}^{\left(  -+\right)  }  &
=\left\vert \left(  \overline{p_{1}}-\Delta p_{1\alpha}^{(-)}\right)  \left(
\overline{p_{2}}-\Delta p_{2\alpha}^{(+)}\right)  \right\vert .
\end{align}
The statistical average can be expressed as%
\begin{align}
\left\langle \Delta p_{1}\Delta p_{2}\right\rangle  &  =\frac{1}{N}%
\sum_{\alpha=1}^{N}\left(  \left(  \Delta p_{1}\Delta p_{2}\right)  _{\alpha
}^{\left(  ++\right)  }+\left(  \Delta p_{1}\Delta p_{2}\right)  _{\alpha
}^{\left(  --\right)  }\right. \nonumber\\
&  +\left.  \left(  \Delta p_{1}\Delta p_{2}\right)  _{\alpha}^{\left(
+-\right)  }+\left(  \Delta p_{1}\Delta p_{2}\right)  _{\alpha}^{\left(
-+\right)  }\right)  .
\end{align}
\begin{figure}[ptbh]
\centering
\fbox{\includegraphics[width=1\linewidth]{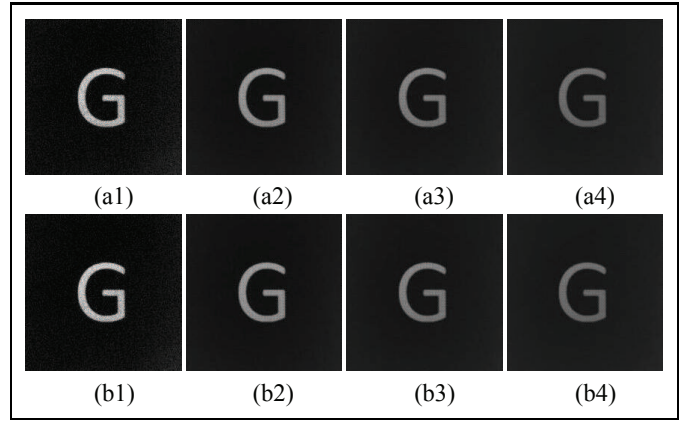}}\caption{Top row: the
images reconstructed by the photon number fluctuation correlation algorithm.
The numbers of measurements are (a1) 10, (a2) 100, (a3) 200, and (a4) 300.
Bottom row: the images reconstructed by the photon number correlation
algorithm. The numbers of measurements are (b1) 10, (b2) 100, (b3) 200, and
(b4) 300.}%
\label{fig:false-color}%
\end{figure}\begin{figure}[ptbh]
\centering
\fbox{\includegraphics[width=1\linewidth]{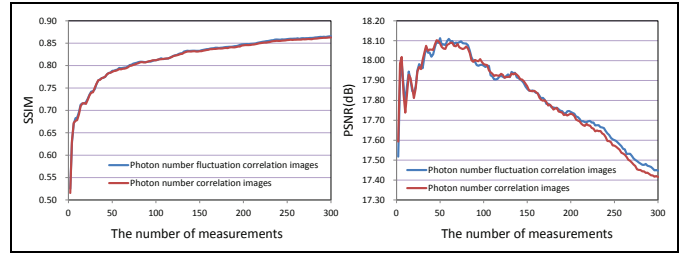}}\caption{ SSIM and PSNR
curves of the images reconstructed by the photon number fluctuation
correlation algorithm (blue) and the photon number correlation algorithm
(red).}%
\label{fig:false-color}%
\end{figure}

Fig. 6 compares the effects of the photon number fluctuation correlation
algorithm and the photon number correlation algorithm. Quantitative
comparisons between these two methods are shown in Fig. 7. If the selective
attenuation of the light spectrum is ignored, the color of the target can be
greatly restored in the defogging image. The simulation results obtained for
color targets are shown in Fig. 8. \begin{figure}[ptbh]
\centering
\fbox{\includegraphics[width=1\linewidth]{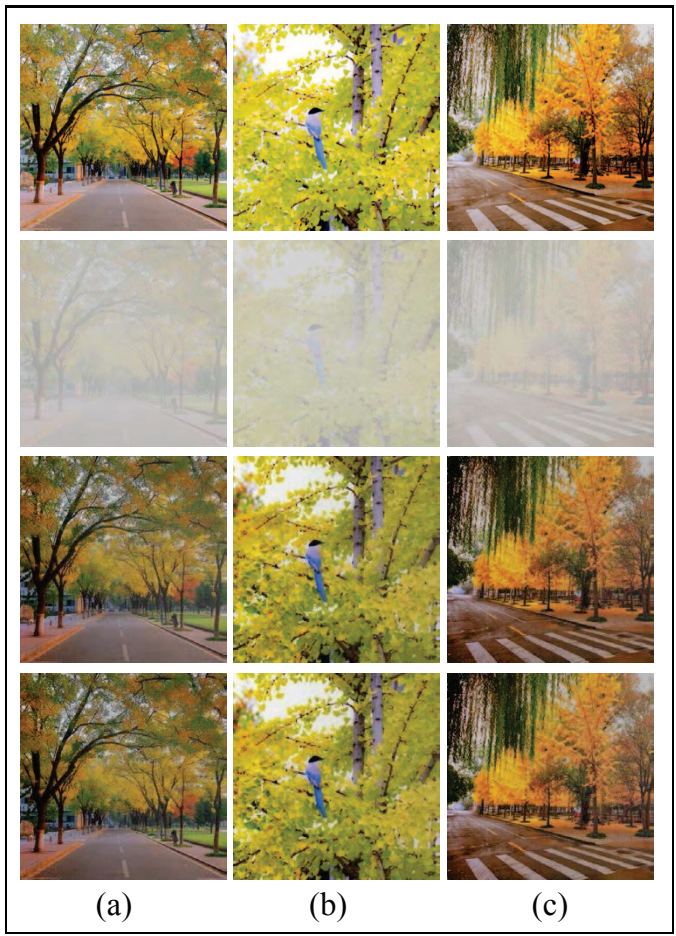}}\caption{First row:
three color targets. Second row: fog images. Third row: the images
reconstructed by the photon number correlation algorithm. Fourth row: the
images reconstructed by the photon number fluctuation correlation algorithm.
The number of measurements is 50. }%
\label{fig:false-color}%
\end{figure}
\begin{figure}[ptbh]
\centering
\fbox{\includegraphics[width=0.5\linewidth]{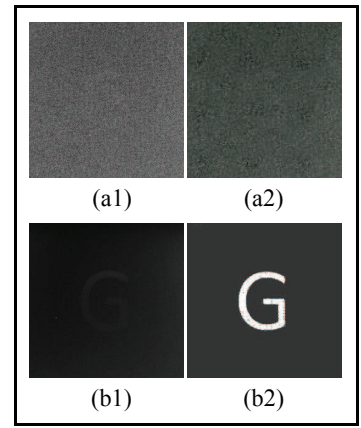}}\caption{(a1)
Indistinguishable fog image. The integration time $\Delta t= 1/30$ second. (a2) The defogging image reconstructed by the cycle
generative adversarial network. (b1) The image reconstructed by the photon number
correlation algorithm with 80 measurements.  Enlarge the images to obtain a more
obvious effect. (b2) The image reconstructed by b1 with a
generative adversarial network.}%
\label{fig:false-color}%
\end{figure}

When the target is indistinguishable, a directly distinguishable image can be
obtained by our methods, although the brightness of the reconstructed image is
low (Fig. 9b1). However, distinguishing these low brightness images is not
difficult for computer vision. Figure 9a2 shows that the target image cannot be
recovered from the indistinguishable image by using a conventional defogging
method with deep learning (for method, see supplemental material). However, a
high-quality image (Fig. 9b2) can be recovered from the reconstructed defogging
images by using deep learning method (for method, see supplemental material).

\newpage
In summary, this article abandons the idea that fog is a scattering medium with
space-time invariance, as assumed by the conventional McCartney model. The correlation
properties of photons interacting with time-variant fog are studied. Our works show
that the scattering photons and the ambient photons have different correlation properties
in the time domain. Thus, a defogging image can be reconstructed by measuring this
difference, which demonstrates that the time-variant nature of fog is a positive
quantum resource. The only requirement is that software must be installed on the
utilized imaging equipment to implement the defogging imaging function without changing any hardware.

This work was supported by the Natural Science Foundation of Shandong Province (ZR2022MF249)
and the National Natural Science Foundation (12274257).

\end{document}